# An Analytical Model for Stepwise Adiabatic Driver Energy Consumption

Eric J. Carlson, Joshua R. Smith

*Abstract*

This paper presents a complete closed-form analytical model for determining the per-cycle energy consumption of stepwise adiabatic drivers used for driving a capacitive load such as a power FET gate. The model takes into account the number of steps used, the stepwise driver tank capacitance, the load capacitance, and the stepwise driver switch resistance and on-time. Model accuracy is compared to that of simulation and models from previous work.

# 1 Introduction

With the developments of the Internet of Things and low-power battery-operated or battery-free devices, numerous techniques have been employed to reduce power consumption of electrical circuits. One such technique is adiabatic charging of capacitive loads that must be charged/discharged periodically, such as large power FETs in DC-DC converters [1], clock lines [2], capacitive touch sensors [3], or others [4], [5]. Inherently, a capacitive load does not consume energy itself when driven with a periodic changing voltage; however, traditional drivers dissipate energy in the process of driving the voltage due to the intrinsic properties of the driver. Adiabatic charging reduces this inherent energy dissipation, and stepwise charging is one such adiabatic technique.

In order to properly optimize the design of a stepwise driver, an accurate analytical model of the stepwise driver energy consumption is needed. Prior works [6], [7], [8] provide formulas to model this energy consumption, but all are inaccurate when voltage steps are not given time to fully settle. Park [3] provides the foundation of a model that does not have such a limitation, but stops short of deriving a complete closed-form formula. This paper presents a complete closed-form analytical formula for calculating stepwise driver energy that is accurate for any amount of step settling. This paper is organized as follows: Section 2 provides an overview of stepwise charging, Section 3 derives an analytical model of the stepwise driver energy consumption, Section 4 compares the model accuracy to that of simulation and to the models of other works, and Section 5 concludes this paper.

# 2 The Stepwise Driver

Stepwise charging involves driving the capacitive load to intermediate voltages in a stepping pattern when transitioning the load voltage $v_{\text{Load}}$ from low to high and then from high to low, as shown in Figure 1. There are variations of how the intermediate voltages are generated [8]. The focus of this work is on the first and simplest variation, introduced in [6], [9]. The intermediate voltages $V_k$ are held by capacitors $C_{\text{Tank},k}$ which naturally become evenly spaced between 0 V and supply voltage $V_{\text{DD}}$ in steady-state, after several step-high/step-low transitions have been performed [6]. In practice, how evenly-spaced the $V_k$ voltages settle to depends on how large the tank capacitance $C_{\text{Tank}}$ is compared to load capacitance $C_{\text{Load}}$ and how much time each step is given to settle.

The number of rising/falling steps in a stepwise driver is defined as $N$. A stepwise driver has $N-1$ $C_{\text{Tank}}$ capacitors, $N$ rising-edge switches $S_R$, and $N$ falling-edge switches $S_F$. In an application such as in [1], the falling-edge switches need to be much lower resistance than the rising-edge switches to enable fast falling transitions. If the resistance of the switches can be equal, then $S_{R,k}$ and $S_{F,N-k}$ (for $k = 1$ to $N-1$) can be one and same device as in [9]. The amount of time that each rising step is given to settle is $T_{\text{SR},k}$ and the amount of time that each falling step is given to settle is $T_{\text{SF},k}$. Figure 1 shows an $N = 5$ stepwise driver. The $v_{\text{Load}}$ voltage at the end of each rising step is defined as $V_{\text{Load},R,k}$, such that $V_{\text{Load},R,1}$ is the voltage at the end of the first step and $V_{\text{Load},R,N}$ is the voltage at the end of the last step. Here, it is assumed that the final step fully settles, such that $V_{\text{Load},R,N} = V_{\text{DD}}$. $V_{\text{Load},R,0} = 0$ V is the pedestal—the starting point of the rising edge. The $v_{\text{Load}}$ voltage at the end of each falling step is defined as $V_{\text{Load},F,k}$, such that $V_{\text{Load},F,0} = V_{\text{Load},R,N} = V_{\text{DD}}$ and $V_{\text{Load},F,1}$ is the first falling step. The final falling step, $V_{\text{Load},F,N}$, is assumed to fully settle, such that $V_{\text{Load},F,N} = V_{\text{Load},R,0} = 0$ V.

From [6], in the most ideal case, where $C_{Tank} \gg C_{Load}$, $S_R$ and $S_F$ are ideal switches, and all steps fully settle, then the energy consumed by the stepwise driver to complete one charge/discharge cycle is

$$E_{\text{Load-Driver}} \approx C_{\text{Load}} \frac{V_{DD}^2}{N}. \tag{1}$$

A single step driver of $N = 1$ is identical to conventional driver that steps directly from 0 V to $V_{DD}$. The next section explains how to calculate the $E_{\text{Load-Driver}}$ energy for the non-ideal case.

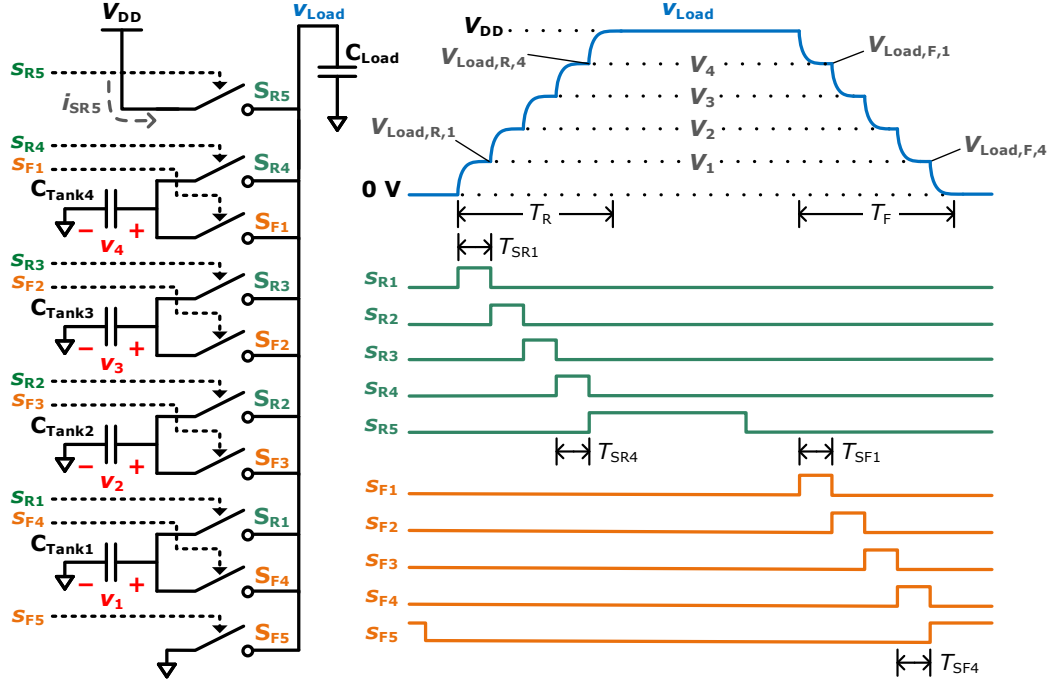

Figure 1. Schematic and waveforms of an $N = 5$ stepwise driver.

## 3 Stepwise Driver Energy Consumption

This section calculates the amount of energy consumed by a stepwise adiabatic driver—like the one shown in Figure 1—based on the following parameters:

- The number of steps: $N$
- The load capacitance: $C_{\text{Load}}$
- The tank capacitance: $C_{\text{Tank}}$
- The rising-edge switch ($S_R$) resistance: $R_{SR}$
- The falling-edge switch ($S_F$) resistance: $R_{SF}$
- The rising-edge switch ($S_R$) on-time: $T_{SR}$
- The falling-edge switch ($S_F$) on-time: $T_{SF}$
- The switch quality factor (defines the energy required to turn a switch $S_R/S_F$ on then off, for a given switch resistance $R_{SR}/R_{SF}$): $\rho$

The stepwise driver energy is divided into two parts. The first part is the energy that is dissipated within the switches of the stepwise driver as it drives the load voltage $v_{Load}$ high and then low for one cycle. This energy will be called $E_{Load\text{-}Driver}$ and is the focus of this paper. The second part is the energy that is consumed by the circuitry that drives the stepwise switches $S_R/S_F$ for each cycle. This energy is discussed in detail in [6] and is briefly discussed at the end of this section.

To simplify the derivation of $E_{Load\text{-}Driver}$, the following assumptions are made:

- All $C_{Tank}$ are equal.
- $T_{SR}$ is the same for every rising step, except for the final step.
- $T_{SF}$ is the same for every falling step, except for the final step.
- $R_{SR}$ is the same for all rising switches $S_R$.
- $R_{SF}$ is the same for all falling switches $S_F$.
- The final rising step is allowed to fully settle to $V_{DD}$.
- The final falling step is allowed to fully settle to 0 V.

The stepwise driver energy $E_{Load\text{-}Driver}$ is the total energy dissipated by the switches in the stepwise driver after one cycle. Svensson [6] approximates this energy by integrating the power consumed by each switch while it is on, but is not fully accurate because it assumes that the tank capacitor voltages $V_k$ are uniformly distributed, which is an approximation. Park [3] proposes to instead find the energy that is drawn from the power supply $V_{DD}$. Since all of the dissipated energy must come from $V_{DD}$, $E_{Load\text{-}Driver}$ must also equal the energy that flows from $V_{DD}$ on the final rising step:

$$E_{Load\text{-}Driver} = \int_{T_{SR,N,on}}^{T_{SR,N,off}} V_{DD} i_{SR,N}, \tag{2}$$

where $T_{SR,N,off} - T_{SR,N,on} = T_{SR,N}$ are the times that switch $S_{R,N}$ turns off (opened) and on (closed), and $i_{SR,N}$ is the current that flows through $S_{R,N}$ while it is conducting. $S_{R,N}$ is the only switch connected to $V_{DD}$. Equation (2) can be re-written in terms of the total charge that flows from $V_{DD}$ through $S_{R,N}$ per cycle:

$$E_{Load\text{-}Driver} = V_{DD} Q_{SR,N}, \tag{3}$$

where $Q_{SR,N}$ is the charge that flows from $V_{DD}$, through the final rising-edge switch $S_{R,N}$, and to $C_{Load}$ on the final rising step, such that:

$$Q_{SR,N} = C_{Load}(V_{Load,R,N} - V_{Load,R,N-1}), \tag{4}$$

where $V_{Load,R,N} = V_{DD}$ is the resting voltage of the final rising step (step $N$) of the load voltage $v_{Load}$. $V_{Load,R,N-1}$ is the resting voltage of the second-to-last rising-edge step and it is the step that connects to the highest $C_{Tank}$ capacitor: $C_{Tank,N-1}$. In the case that the final rising step is given enough time to fully settle to the final value ($V_{DD}$) the energy becomes

$$E_{Load\text{-}Driver} = C_{Load} V_{DD}(V_{DD} - V_{Load,R,N-1}). \tag{5}$$

To find $E_{Load\text{-}Driver}$, $V_{Load,R,N-1}$ must be determined. Park [3] provides some of the equations required to solve for $V_{Load,R,N-1}$, but stops short of deriving a final solution. The first step in finding $V_{Load,R,N-1}$ is to define terms $r$ (for rising-edge) and $f$ (for falling-edge), which represent the percentage that each step reaches the average of the voltage of the $C_{\text{Tank},k}$ capacitor that is being stepped into $V_k$, defined as:

$$r_k := \frac{V_{\text{Load},R,k} - V_{\text{Load},R,k-1}}{V_k - V_{\text{Load},R,k-1}}, \tag{6}$$

$$f_k := \frac{V_{\text{Load},F,k} - V_{\text{Load},F,k-1}}{V_{N-k} - V_{\text{Load},F,k-1}}, \tag{7}$$

for $0 < k < N$ and such that $0 < r < 1$ and $0 < f < 1$. The analysis that follows will assume that $r$ and $f$ are equal for all steps $k$ (except for the final rising step $r_N$ and the final falling step $f_N$, which equal 1):

$$r = r_k, \tag{8}$$

$$f = f_k. \tag{9}$$

For an ideal stepwise driver (where $C_{\text{Tank}} \gg C_{\text{Load}}$ and each step is given time to fully settle), $r = f = 1$. There are two reasons why $r$ and $f$ will be less than one. First, due to the RC settling time constant $\tau_R/\tau_F$, the switch $S_R/S_F$ might open before $v_{\text{Load}}$ fully settles. Second, charge redistribution between $C_{\text{Load}}$ and $C_{\text{Tank}}$ will cause the voltage $v_k$ across capacitor $C_{\text{Tank},k}$ to have ripple $\Delta V_k$. Even if the $v_{\text{Load}}$ step fully settles such that $v_{\text{Load}} = v_k$, at the end of the step $v_k$ will be less than the average $C_{\text{Tank},k}$ voltage $V_k$ due to this ripple. Figure 2 illustrates the charging behavior of one step $k$, while Figure 3 illustrates the discharging behavior for one falling step $N - k$. The dashed lines represent what the settled voltages would be if switches $S_{R,k}$ ($S_{F,N-k}$) never opened and $S_{R,k+1}$ ($S_{F,N-k+1}$) never closed.

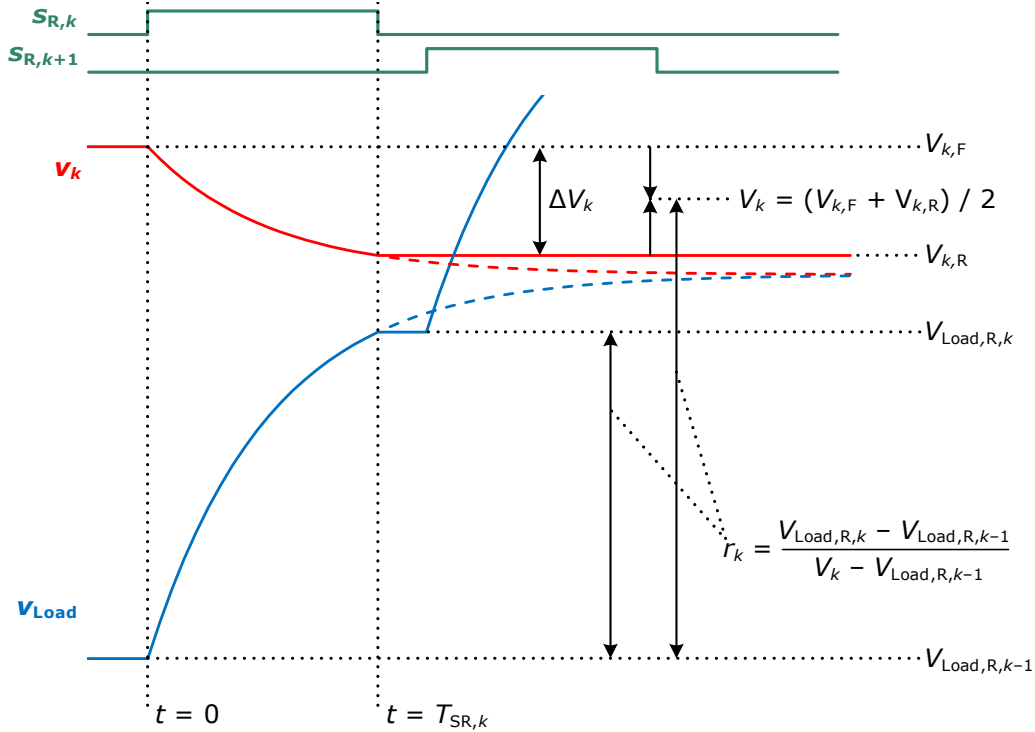

Figure 2. Waveforms of the load voltage $v_{\text{Load}}$ and $C_{\text{Tank},k}$ capacitor voltage $v_k$, for one rising step $k$.

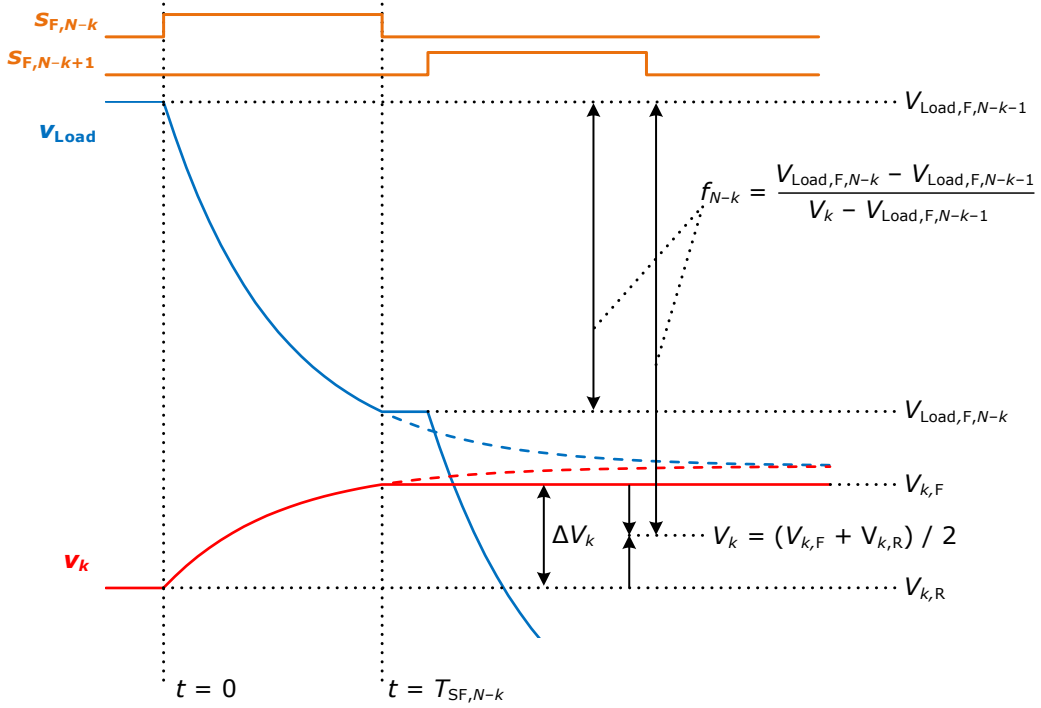

Figure 3. Waveforms of the load voltage $v_{Load}$ and $C_{Tank,k}$ capacitor voltage $v_k$, for one falling step $N-k$.

The voltage across capacitor $C_{Tank,k}$ is defined as $v_k$ for the continuous time voltage, and $V_k$ for the average of $v_k$. However, it is very important to recognize that $V_k$ is not the average of $v_k$ across time. Instead, it is the average of two discrete voltages: the final voltage after rising step $k$ completes, $V_{k,R}$, and the final voltage after falling step $N-k$ completes, $V_{k,F}$:

$$\Delta V_k := V_{k,F} - V_{k,R}, \tag{10}$$

$$V_k := \frac{V_{k,F} + V_{k,R}}{2}. \tag{11}$$

For rising steps, the load voltage at the beginning of a step ($t = 0$), when the switch $S_{R,k}$ first closes is equal to the final voltage of the previous step, right after switch $S_{R,k-1}$ opens:

$$v_{Load,(t=0)} = V_{Load,R,k-1}. \tag{12}$$

The $C_{Tank,k}$ voltage that is connected to the load through $S_{R,k}$ at $t = 0$ is equal to the voltage level at the end of the last falling step, when $S_{F,N-k}$ opened:

$$v_{k,(t=0)} = V_{k,F}. \tag{13}$$

The load voltage $v_{Load}$ and $C_{Tank,k}$ voltage $v_k$ follow an RC-charging waveform such that

$$v_{Load} = \left(V_{k,F} - V_{Load,R,k-1}\right)\left(\frac{C_{Series}}{C_{Load}}\left(1 - e^{-\frac{t}{\tau_R}}\right)\right) + V_{Load,R,k-1}, \tag{14}$$

$$v_k = \left(V_{k,\text{F}} - V_{\text{Load,R},k-1}\right)\left(e^{-\frac{t}{\tau_\text{R}}} + \frac{C_{\text{Series}}}{C_{\text{Load}}}\left(1 - e^{-\frac{t}{\tau_\text{R}}}\right)\right) + V_{\text{Load,R},k-1}, \tag{15}$$

where

$$C_{\text{Series}} = \frac{C_{\text{Tank}} C_{\text{Load}}}{C_{\text{Tank}} + C_{\text{Load}}}, \tag{16}$$

$$\tau_\text{R} = C_{\text{Series}} R_{\text{SR}}, \tag{17}$$

and $R_{\text{SR}}$ is the resistance of the rising-edge switches $S_\text{R}$.

The $v_{\text{Load}}$ and $v_k$ voltages at the time that the switch $S_{\text{R},k}$ opens ($t = T_{\text{SR}}$) become

$$V_{\text{Load,R},k} = v_{\text{Load},(t=T_{\text{SR}})} = \left(V_{k,\text{F}} - V_{\text{Load,R},k-1}\right)\left(\frac{C_{\text{Series}}}{C_{\text{Load}}}\left(1 - e^{-\frac{T_{\text{SR}}}{\tau_\text{R}}}\right)\right) + V_{\text{Load,R},k-1}. \tag{18}$$

$$V_{k,\text{R}} = v_{k,(t=T_{\text{SR}})} = \left(V_{k,\text{F}} - V_{\text{Load,R},k-1}\right)\left(e^{-\frac{T_{\text{SR}}}{\tau_\text{R}}} + \frac{C_{\text{Series}}}{C_{\text{Load}}}\left(1 - e^{-\frac{T_{\text{SR}}}{\tau_\text{R}}}\right)\right) + V_{\text{Load,R},k-1}. \tag{19}$$

Combining (11) and (19) gives

$$V_k = \frac{V_{k,\text{F}} + \left(V_{k,\text{F}} - V_{\text{Load,R},k-1}\right)\left(e^{-\frac{T_{\text{SR}}}{\tau_\text{R}}} + \frac{C_{\text{Series}}}{C_{\text{Load}}}\left(1 - e^{-\frac{T_{\text{SR}}}{\tau_\text{R}}}\right)\right) + V_{\text{Load,R},k-1}}{2}. \tag{20}$$

Combining (6) and (8) with (20) gives

$$r = \frac{V_{\text{Load,R},k} - V_{\text{Load,R},k-1}}{\frac{V_{k,\text{F}} + \left(V_{k,\text{F}} - V_{\text{Load,R},k-1}\right)\left(e^{-\frac{T_{\text{SR}}}{\tau_\text{R}}} + \frac{C_{\text{Series}}}{C_{\text{Load}}}\left(1 - e^{-\frac{T_{\text{SR}}}{\tau_\text{R}}}\right)\right) + V_{\text{Load,R},k-1}}{2} - V_{\text{Load,R},k-1}}, \tag{21}$$

which can be simplified to

$$r = 2\frac{V_{\text{Load,R},k} - V_{\text{Load,R},k-1}}{\left(V_{k,\text{F}} - V_{\text{Load,R},k-1}\right)\left(1 + e^{-\frac{T_{\text{SR}}}{\tau_\text{R}}} + \frac{C_{\text{Series}}}{C_{\text{Load}}}\left(1 - e^{-\frac{T_{\text{SR}}}{\tau_\text{R}}}\right)\right)}. \tag{22}$$

Substituting (18) into (22) produces

$$r = 2\frac{\left(V_{k,\text{F}} - V_{\text{Load,R},k-1}\right)\left(\frac{C_{\text{Series}}}{C_{\text{Load}}}\left(1 - e^{-\frac{T_{\text{SR}}}{\tau_\text{R}}}\right)\right) + V_{\text{Load,R},k-1} - V_{\text{Load,R},k-1}}{\left(V_{k,\text{F}} - V_{\text{Load,R},k-1}\right)\left(1 + e^{-\frac{T_{\text{SR}}}{\tau_\text{R}}} + \frac{C_{\text{Series}}}{C_{\text{Load}}}\left(1 - e^{-\frac{T_{\text{SR}}}{\tau_\text{R}}}\right)\right)}, \tag{23}$$

which can be simplified to

$$r = \frac{2C_{\text{Series}}\left(1 - e^{-\frac{T_{\text{SR}}}{\tau_{\text{R}}}}\right)}{C_{\text{Load}}\left(1 + e^{-\frac{T_{\text{SR}}}{\tau_{\text{R}}}}\right) + C_{\text{Series}}\left(1 - e^{-\frac{T_{\text{SR}}}{\tau_{\text{R}}}}\right)}, \quad (24)$$

and further simplified to

$$r = \frac{2C_{\text{Series}}}{C_{\text{Series}} + C_{\text{Load}}\dfrac{1 + e^{-\frac{T_{\text{SR}}}{\tau_{\text{R}}}}}{1 - e^{-\frac{T_{\text{SR}}}{\tau_{\text{R}}}}}}. \quad (25)$$

Applying a trigonometric identity, (25) can be written as

$$r = \frac{2C_{\text{Series}}}{C_{\text{Series}} + C_{\text{Load}} \coth\left(\frac{T_{\text{SR}}}{2\tau_{\text{R}}}\right)}. \quad (26)$$

From symmetry, the falling-edge step term $f$ can be reached in the same way:

$$f = \frac{2C_{\text{Series}}}{C_{\text{Series}} + C_{\text{Load}} \coth\left(\frac{T_{\text{SF}}}{2\tau_{\text{F}}}\right)}, \quad (27)$$

where

$$\tau_{\text{F}} = C_{\text{Series}} R_{\text{SF}}. \quad (28)$$

Now that the $r$ and $f$ terms have been defined and derived, the next step is to use those terms to set up a system of equations to determine the value of $v_{\text{Load}}$ at the end of each rising step $V_{\text{Load,R},k}$ as a function of $r$, and $V_k$. In the case of an $N = 5$ step driver, $v_{\text{Load}}$ at the end of each step is:

$$\begin{aligned}
\text{pedistal: } & V_{\text{Load,R},0} = 0\text{ V} \\
\text{step 1: } & V_{\text{Load,R},1} = rV_1 \\
\text{step 2: } & V_{\text{Load,R},2} = V_{\text{Load,R},1} + r(V_2 - V_{\text{Load,R},1}) \\
\text{step 3: } & V_{\text{Load,R},3} = V_{\text{Load,R},2} + r(V_3 - V_{\text{Load,R},2}) \\
\text{step 4: } & V_{\text{Load,R},4} = V_{\text{Load,R},3} + r(V_4 - V_{\text{Load,R},3}) \\
\text{step 5: } & V_{\text{Load,R},5} = V_{\text{DD}}
\end{aligned} \quad (29)$$

In general, for arbitrary $N$:

$$V_{\text{Load,R},k} = \begin{cases} 0, & k = 0 \\ V_{\text{Load,R},k-1} + r(V_k - V_{\text{Load,R},k-1}), & 0 < k < N \\ V_{\text{DD}}, & k = N \end{cases} \quad (30)$$

This can be re-written as:

$$\begin{aligned}
&\text{pedistal: } V_{\text{Load,R},0} = 0 \\
&\text{step 1: } V_{\text{Load,R},1} = rV_1 \\
&\text{step 2: } V_{\text{Load,R},2} = rV_1 + r(V_2 - rV_1) \\
&\text{step 3: } V_{\text{Load,R},3} = rV_1 + r(V_2 - rV_1) + r\big(V_3 - rV_1 + r(V_2 - rV_1)\big) \\
&\text{step 4: } V_{\text{Load,R},4} = rV_1 + r(V_2 - rV_1) + r\big(V_3 - rV_1 + r(V_2 - rV_1)\big) + \\
&\qquad r\Big(V_4 - rV_1 + r(V_2 - rV_1) + r\big(V_3 - rV_1 + r(V_2 - rV_1)\big)\Big) \\
&\text{step 5: } V_{\text{Load,R},5} = V_{\text{DD}}
\end{aligned} \qquad (31)$$

which can be re-written as:

$$\begin{aligned}
&\text{pedistal: } V_{\text{Load,R},0} = 0 \\
&\text{step 1: } V_{\text{Load,R},1} = rV_1 \\
&\text{step 2: } V_{\text{Load,R},2} = r(1-r)V_1 + rV_2 \\
&\text{step 3: } V_{\text{Load,R},3} = r(1-r)^2 V_1 + r(1-r)V_2 + rV_3 \\
&\text{step 4: } V_{\text{Load,R},4} = r(1-r)^3 V_1 + r(1-r)^2 V_2 + r(1-r)V_3 + rV_4 \\
&\text{step 5: } V_{\text{Load,R},5} = V_{\text{DD}}
\end{aligned} \qquad (32)$$

In general, for arbitrary $N$:

$$V_{\text{Load,R},k} = \begin{cases} 0, & k = 0 \\ \sum_{i=1}^{k} r(1-r)^{k-i} V_i, & 0 < k < N, \\ V_{\text{DD}}, & k = N \end{cases} \qquad (33)$$

Where $V_i$ in this case is $C_{\text{Tank},k}$ voltage $V_k$, but with $k$ replaced with $i$ for the summation.

Next, the load voltage at the end of each falling step $V_{\text{Load,F},k}$ is determined as a function of $f$ and $V_k$. For the falling-edge steps, the $k^{\text{th}}$ step of $v_{\text{Load}}$ does not distribute charge to $C_{\text{Tank},k}$ ($V_k$) as it does with the rising steps. Instead, on step $k$, $S_{\text{F},k}$ closes and discharges $v_{\text{Load}}$ into $C_{\text{Tank},N-k}$ ($V_{N-k}$). This is the reverse of the rising steps. For an $N = 4$ step driver:

$$\begin{aligned}
&\text{plateau: } V_{\text{Load,F},0} = V_{\text{DD}} \\
&\text{step 1: } V_{\text{Load,F},1} = V_{\text{Load,F},0} - f(V_{\text{Load,F},0} - V_3) \\
&\text{step 2: } V_{\text{Load,F},2} = V_{\text{Load,F},1} - f(V_{\text{Load,F},1} - V_2). \\
&\text{step 3: } V_{\text{Load,F},3} = V_{\text{Load,F},2} - f(V_{\text{Load,F},2} - V_1) \\
&\text{step 4: } V_{\text{Load,F},4} = 0 \text{ V}
\end{aligned} \qquad (34)$$

In general, for arbitrary $N$:

$$V_{\text{Load,F},k} = \begin{cases} V_{\text{DD}}, & k = 0 \\ V_{\text{Load,F},k-1} - f(V_{\text{Load,F},k-1} - V_{N-k}), & 0 < k < N. \\ 0, & k = N \end{cases} \qquad (35)$$

This can be rewritten as:

$$\begin{aligned}
\text{plateau: } & V_{\text{Load,F},0} = V_{\text{DD}} \\
\text{step 1: } & V_{\text{Load,F},1} = V_{\text{Load,F},0} - f(V_{\text{Load,F},0} - V_3) \\
\text{step 2: } & V_{\text{Load,F},2} = V_{\text{Load,F},1} - f(V_{\text{Load,F},1} - V_2), \\
\text{step 3: } & V_{\text{Load,F},3} = V_{\text{Load,F},2} - f(V_{\text{Load,F},2} - V_1) \\
\text{step 4: } & V_{\text{Load,F},4} = 0 \text{ V}
\end{aligned} \qquad (36)$$

which can be expanded to become

$$\begin{aligned}
\text{plateau: } & V_{\text{Load,F},0} = V_{\text{DD}} \\
\text{step 1: } & V_{\text{Load,F},1} = V_{\text{DD}} - f(V_{\text{DD}} - V_3) \\
\text{step 2: } & V_{\text{Load,F},2} = V_{\text{DD}} - f(V_{\text{DD}} - V_3) - f(V_{\text{DD}} - f(V_{\text{DD}} - V_3) - V_2) \\
\text{step 3: } & V_{\text{Load,F},3} = V_{\text{DD}} - f(V_{\text{DD}} - V_3) - f(V_{\text{DD}} - f(V_{\text{DD}} - V_3) - V_2) - \\
& \qquad f(V_{\text{DD}} - f(V_{\text{DD}} - V_3) - f(V_{\text{DD}} - f(V_{\text{DD}} - V_3) - V_2) - V_1) \\
\text{step 4: } & V_{\text{Load,F},4} = 0 \text{ V}
\end{aligned} \qquad (37)$$

This can be simplified to

$$\begin{aligned}
\text{plateau: } & V_{\text{Load,F},0} = V_{\text{DD}} \\
\text{step 1: } & V_{\text{Load,F},1} = (1-f)V_{\text{DD}} + fV_3 \\
\text{step 2: } & V_{\text{Load,F},2} = (1-f)^2 V_{\text{DD}} + f(1-f)V_3 + fV_2 \\
\text{step 3: } & V_{\text{Load,F},3} = (1-f)^3 V_{\text{DD}} + f(1-f)^2 V_3 + f(1-f)V_2 + fV_1 \\
\text{step 4: } & V_{\text{Load,F},4} = 0 \text{ V}
\end{aligned} \qquad (38)$$

In general, for arbitrary $N$:

$$V_{\text{Load,F},k} = \begin{cases} V_{\text{DD}}, & k = 0 \\ V_{\text{DD}}(1-f)^k + \sum_{i=0}^{k-1} f(1-f)^i V_{N-k+i}, & 0 < k < N \\ 0, & k = N \end{cases} \qquad (39)$$

For N = 5:

$$\begin{aligned}
\text{plateau: } & V_{\text{Load,F},0} = V_{\text{DD}} \\
\text{step 1: } & V_{\text{Load,F},1} = (1-f)V_{\text{DD}} + fV_4 \\
\text{step 2: } & V_{\text{Load,F},2} = (1-f)^2 V_{\text{DD}} + f(1-f)V_4 + fV_3 \\
\text{step 3: } & V_{\text{Load,F},3} = (1-f)^3 V_{\text{DD}} + f(1-f)^2 V_4 + f(1-f)V_3 + fV_2 \\
\text{step 4: } & V_{\text{Load,F},4} = (1-f)^4 V_{\text{DD}} + f(1-f)^3 V_4 + f(1-f)^2 V_3 + f(1-f)V_2 + fV_1 \\
\text{step 5: } & V_{\text{Load,F},5} = 0 \text{ V}
\end{aligned} \qquad (40)$$

Sets of equations (33) and (39) leave 3($N$−1) unknowns: $V_k$, $V_{\text{Load,R},k}$, $V_{\text{Load,F},k}$ (for $k = 1$ to $N - 1$), with only 2($N$−1) equations. The final set of equations needed to solve for these variables is derived from the fact that, when in steady-state, the charge leaving a given $C_{\text{Tank},k}$ capacitor on a rising step ($-\Delta Q_{k,\text{R}}$) must equal the charge entering the capacitor on the falling step ($\Delta Q_{k,\text{F}}$), such that:

$$-\Delta Q_{k,\text{R}} = \Delta Q_{k,\text{F}}. \qquad (41)$$

This means that the voltage step size when connecting the load capacitor $C_{\text{Load}}$ to capacitor $C_{\text{Tank},k}$ on the rising edge (when $S_{\text{R},k}$ is closed) must equal the voltage step size when connecting $C_{\text{Load}}$ to $C_{\text{Tank},k}$ on the falling edge ($S_{\text{F},N-k}$ is closed):

$$V_{\text{Load,R},k} - V_{\text{Load,R},k-1} = V_{\text{Load,F},N-k-1} - V_{\text{Load,F},N-k}. \tag{42}$$

This can be rewritten as

$$V_{\text{Load,R},k} + V_{\text{Load,F},N-k} = V_{\text{Load,R},k-1} + V_{\text{Load,F},N-k-1}. \tag{43}$$

For $N = 5$, (43) becomes

$$\begin{aligned} k &= 1: V_{\text{Load,R},1} + V_{\text{Load,F},4} = V_{\text{Load,R},0} + V_{\text{Load,F},3} \\ k &= 2: V_{\text{Load,R},2} + V_{\text{Load,F},3} = V_{\text{Load,R},1} + V_{\text{Load,F},2} \\ k &= 3: V_{\text{Load,R},3} + V_{\text{Load,F},2} = V_{\text{Load,R},2} + V_{\text{Load,F},1} \\ k &= 4: V_{\text{Load,R},4} + V_{\text{Load,F},1} = V_{\text{Load,R},3} + V_{\text{Load,F},0} \end{aligned} \tag{44}$$

Substituting (33) and (39) into (44) yields

$$\begin{aligned} k = 1: \ & rV_1 + (1-f)^4 V_{\text{DD}} + f(1-f)^3 V_4 + f(1-f)^2 V_3 + f(1-f)V_2 + fV_1 \\ &= 0\,\text{V} + (1-f)^3 V_{\text{DD}} + f(1-f)^2 V_4 + f(1-f)V_3 + fV_2 \\ k = 2: \ & r(1-r)V_1 + (1-f)^3 V_{\text{DD}} + f(1-f)^2 V_4 + f(1-f)V_3 + fV_2 \\ &= rV_1 + (1-f)^2 V_{\text{DD}} + f(1-f)V_4 + fV_3 \\ k = 3: \ & r(1-r)^2 V_1 + r(1-r)V_2 + rV_3 + (1-f)^2 V_{\text{DD}} + f(1-f)V_4 + fV_3 \\ &= r(1-r)V_1 + rV_2 + (1-f)V_{\text{DD}} + fV_4 \\ k = 4: \ & r(1-r)^3 V_1 + r(1-r)^2 V_2 + r(1-r)V_3 + rV_4 + (1-f)V_{\text{DD}} + fV_4 \\ &= r(1-r)^2 V_1 + r(1-r)V_2 + rV_3 + V_{\text{DD}} \end{aligned} \tag{45}$$

which can be simplified to

$$\begin{aligned} k = 1: \ & -r^2(1-r)^2 V_1 - r^2(1-r)V_2 - r^2 V_3 + (r+f)V_4 = fV_{\text{DD}} \\ k = 2: \ & -r^2(1-r)V_1 - r^2 V_2 + (r+f)V_3 - f^2 V_4 = f(1-f)V_{\text{DD}} \\ k = 3: \ & -r^2 V_1 + (r+f)V_2 - f^2 V_3 - f^2(1-f)V_4 = f(1-f)^2 V_{\text{DD}} \\ k = 4: \ & (r+f)V_1 - f^2 V_2 - f^2(1-f)V_3 - f^2(1-f)^2 V_4 = f(1-f)^3 V_{\text{DD}} \end{aligned} \tag{46}$$

This can be formed into a matrix equation:

$$\mathbf{AV} = \mathbf{B}, \tag{47}$$

$$\begin{bmatrix} -r^2(1-r)^2 & -r^2(1-r) & -r^2 & r+f \\ -r^2(1-r) & -r^2 & r+f & -f^2 \\ -r^2 & r+f & -f^2 & -f^2(1-f) \\ r+f & -f^2 & -f^2(1-f) & -f^2(1-f)^2 \end{bmatrix} \begin{bmatrix} V_1 \\ V_2 \\ V_3 \\ V_4 \end{bmatrix} = \begin{bmatrix} fV_{\text{DD}} \\ f(1-f)V_{\text{DD}} \\ f(1-f)^2 V_{\text{DD}} \\ f(1-f)^3 V_{\text{DD}} \end{bmatrix}. \tag{48}$$

These matrices form a clear pattern, which is more visible when including all (1-*r*) and (1-*f*) exponents:

$$\begin{bmatrix} -r^2(1-r)^2 & -r^2(1-r)^1 & -r^2(1-r)^0 & r+f \\ -r^2(1-r)^1 & -r^2(1-r)^0 & r+f & -f^2(1-f)^0 \\ -r^2(1-r)^0 & r+f & -f^2(1-f)^0 & -f^2(1-f)^1 \\ r+f & -f^2(1-f)^0 & -f^2(1-f)^1 & -f^2(1-f)^2 \end{bmatrix} \begin{bmatrix} V_1 \\ V_2 \\ V_3 \\ V_4 \end{bmatrix} = \begin{bmatrix} f(1-f)^0 V_{DD} \\ f(1-f)^1 V_{DD} \\ f(1-f)^2 V_{DD} \\ f(1-f)^3 V_{DD} \end{bmatrix}. \quad (49)$$

The pattern can be extrapolated to any value for $N > 1$.

For $N = 2$:

$$(r+f)V_1 = fV_{DD}. \quad (50)$$

For $N = 6$:

$$\begin{bmatrix} r^2(1-r)^3 & -r^2(1-r)^2 & -r^2(1-r)^1 & -r^2(1-r)^0 & r+f \\ -r^2(1-r)^2 & -r^2(1-r)^1 & -r^2(1-r)^0 & r+f & -f^2(1-f)^0 \\ -r^2(1-r)^1 & -r^2(1-r)^0 & r+f & -f^2(1-f)^0 & -f^2(1-f)^1 \\ -r^2(1-r)^0 & r+f & -f^2(1-f)^0 & -f^2(1-f)^1 & -f^2(1-f)^2 \\ r+f & -f^2(1-f)^0 & -f^2(1-f)^1 & -f^2(1-f)^2 & -f^2(1-f)^3 \end{bmatrix} \begin{bmatrix} V_1 \\ V_2 \\ V_3 \\ V_4 \\ V_5 \end{bmatrix}$$
$$= \begin{bmatrix} f(1-f)^0 V_{DD} \\ f(1-f)^1 V_{DD} \\ f(1-f)^2 V_{DD} \\ f(1-f)^3 V_{DD} \\ f(1-f)^4 V_{DD} \end{bmatrix}. \quad (51)$$

With this, all $V_k$ voltages can be solved for. Once all $V_k$ voltages are known, $V_{\text{Load,R},N-1}$ can be solved for with (33). After $V_{\text{Load,R},N-1}$ is known, $E_{\text{Load-Driver}}$ can be determined from (5).

The following pseudocode builds the **A** matrix and **B** vector:

```
A = zero array of size (N-1) by (N-1)
B = zero array of size (N-1) by 1
for i in [1,2,3,...,(N-1)] :
    B[i] = f * (1-f)**i * VDD
    for j in [1,2,3,...,(N-1)] :
        if (i+j == N-2) :
            A[i,j] = r+f
        else if (i+j < N-2) :
            A[i,j] = -(r**2) * (1 - r)**(N-3-i-j)
        else if (i+j > N-2) :
            A[i,j] = -(f**2) * (1-f)**(i+j-(N-1))
```

In the pseudocode above, **\*\*** symbolizes "to the power of".

Equation (5) does not consider the energy required to drive the switches $S_R/S_F$ on and off. The total energy required to drive the load is

$$E_{\text{Driver,Total}} = E_{\text{Load-Driver}} + E_{\text{Switch-Driver}}, \qquad (52)$$

where $E_{\text{Switch-Driver}}$ is the combined total of the energy consumed to drive all of the switches $S_R/S_F$ in the stepwise driver, such that

$$E_{\text{Switch-Driver}} = \sum_{k=1}^{N} E_{\text{SR},k} + \sum_{k=1}^{N} E_{\text{SF},k}. \qquad (53)$$

If the switches $S_R/S_F$ are implemented as MOSFET devices, the energy required to drive each individual switch is inversely proportional to the on-state (closed) resistance of the switch:

$$E_{\text{SR},k} = \frac{\rho_{R,k}}{R_{\text{SR},k}}, \qquad E_{\text{SF},k} = \frac{\rho_{F,k}}{R_{\text{SF},k}}. \qquad (54)$$

Terms $\rho_{R,k}$ and $\rho_{F,k}$ are the quality factor of the MOSFET switch and symbolizes the energy required to turn a switch of a given resistance on then off. If one assumes that all $S_R$ are identical and all $S_F$ are identical, then (53) simplifies to

$$E_{\text{Switch-Driver}} = N E_{\text{SR}} + N E_{\text{SF}}. \qquad (55)$$

The remaining discussion focuses on $E_{\text{Load-Driver}}$.

## 4 Results

This section investigates the accuracy of the stepwise driver model for $E_{\text{Load-Driver}}$ derived in the previous section by comparing it to the results of a circuit simulator. This section also compares the accuracy with that of models from previous work.

### A. Comparison to Simulated Data

To validate the model, the calculated $E_{\text{Load-Driver}}$ was compared to results from a circuit simulator. In the simulation, switches $S_R/S_F$ were modeled as ideal switches with a series resistance of $R_{\text{SR}} = R_{\text{SF}}$. All capacitors were modeled as ideal capacitors. Simulations were run for a sufficient number of step-up/step-down cycles such that all $V_k$ voltages reached their steady-state values. $E_{\text{Load-Driver}}$ was measured in simulation using (2). $E_{\text{Load-Driver}}$ was compared for combinations of $N = 4$ and $N = 9$, as well as $C_{\text{Tank}} = C_{\text{Load}}$ and $C_{\text{Tank}} = 4C_{\text{Load}}$. Switch on-time was swept such that $T_{\text{SR}} = 0$ through $6\tau$. Here, $\tau = \tau_R = \tau_F$. Figure 6 compares calculated results (dotted lines) with simulated results (solid lines) for the case that $T_{\text{SF}} = T_{\text{SR}}$. Figure 7 compares calculated results with simulated results for the case that $T_{\text{SF}} = 2T_{\text{SR}}$. The $E_{\text{Load-Driver}}$ energy is normalized to that of a conventional ($N = 1$) driver. The $E_{\text{Load-Driver}}$ energy calculated using the proposed analytical model closely matches the results from the circuit simulator.

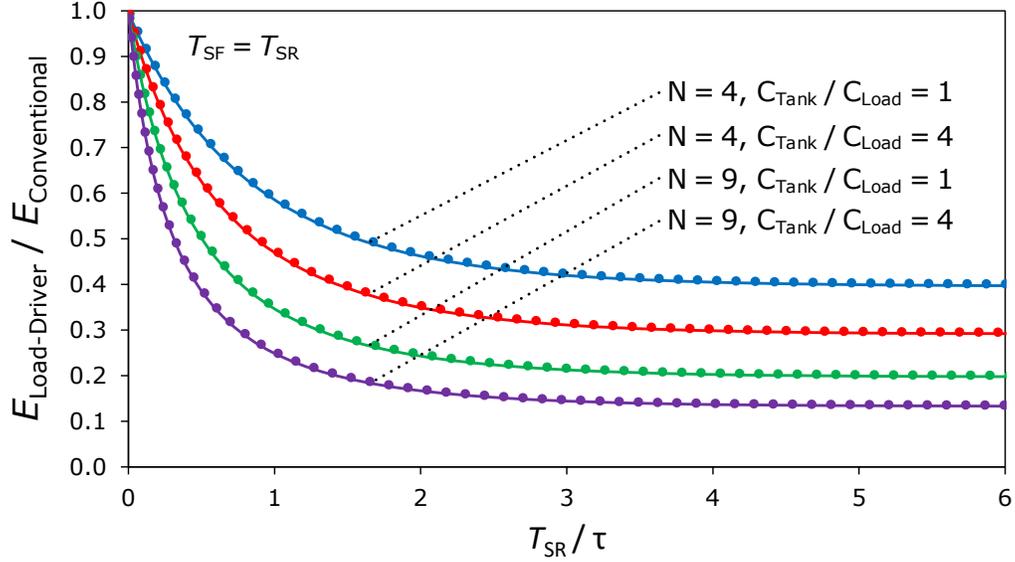

Figure 4. Comparing the proposed stepwise driver model results (dotted line) with results from a circuit simulator (solid line) for various $N$, $C_{Tank}$, and $T_{SR}$ values. Stepwise driver energy is normalized to that of a conventional driver ($N = 1$). Switch $S_R$ on-time $T_{SR}$ is normalized to the RC settling time constant, $\tau = \tau_R = \tau_F$. Switch $S_F$ on-time equals the $S_R$ on-time: $T_{SF} = T_{SR}$.

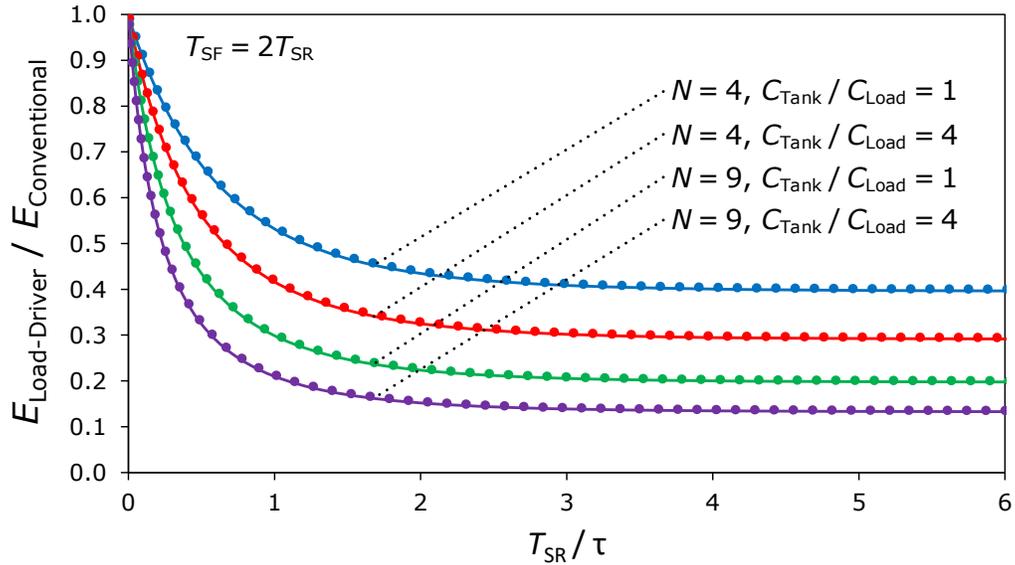

Figure 5. Comparing the proposed stepwise driver model results (dotted line) with results from a circuit simulator (solid line) for various $N$, $C_{Tank}$, and $T_{SR}$ values. Stepwise driver energy is normalized to that of a conventional driver ($N = 1$). Switch $S_R$ on-time $T_{SR}$ is normalized to the RC settling time constant, $\tau = \tau_R = \tau_F$. The $S_F$ on-time is double the $S_R$ on-time: $T_{SF} = 2T_{SR}$.

## B. Comparison to Prior Work

Svensson [6] introduced stepwise charging and provides an approximation of the energy consumed by the stepwise driver based on the number of steps $N$ and the amount of time that each step is given to settle. From [6]:

$$E_{\text{Driver,Total[6]}} = \left(\frac{1}{N}\coth\left(\frac{m}{2}\right) + 2N^2 m \frac{\bar{\rho}}{T}\right) C_{\text{Load}} V_{\text{DD}}^2. \tag{56}$$

In [6], $m$ is the number of time constants that the stepwise driver switches are kept closed. In the terminology of this paper

$$m := \frac{T_{\text{SR}}}{\tau_R}, \tag{57}$$

assuming that $T_{\text{SR}} = T_{\text{SF}}$ and $\tau_R = \tau_F$. Svensson [6] defines $\bar{\rho}$ as the gate capacitance of a stepwise driver switch for a given resistance. Setting $\bar{\rho} = 0$ gives the [6] analogue of $E_{\text{Load-Driver}}$:

$$E_{\text{Load-Driver[6]}} = \frac{1}{N}\coth\left(\frac{T_{\text{SR}}}{2\tau_R}\right) C_{\text{Load}} V_{\text{DD}}^2. \tag{58}$$

Equation (58) from [6] does not take into account finite $C_{\text{Tank}}$. Dancy [7] provides the energy dissipated in the stepwise driver switches for half a charging cycle (only charging the load, not discharging the load) as

$$E_{\text{diss[7]}} = \frac{V_{\text{DD}}^2 C_{\text{Load}}(C_{\text{Tank}} + C_{\text{Load}})}{2(C_{\text{Load}} + NC_{\text{Tank}})}. \tag{59}$$

This is only half of the total energy dissipated by the stepwise driver per complete charge/discharge cycle. It must be doubled to find the total energy:

$$E_{\text{Load-Driver[7]}} = \frac{V_{\text{DD}}^2 C_{\text{Load}}(C_{\text{Tank}} + C_{\text{Load}})}{C_{\text{Load}} + NC_{\text{Tank}}}. \tag{60}$$

This approximation takes into account finite $C_{\text{Tank}}$, but assumes each step fully settles ($T_{\text{SR}} \gg \tau_R$, $T_{\text{SF}} \gg \tau_F$). Although not explicitly stated in [7], it can be deduced that adding the settling-time-inclusive model from [6] into the $C_{\text{Tank}}$-inclusive model in [7] will produce

$$E_{\text{Load-Driver[6][7]}} = \frac{V_{\text{DD}}^2 C_{\text{Load}}(C_{\text{Tank}} + C_{\text{Load}})}{C_{\text{Load}} + \dfrac{N}{\coth\left(\dfrac{T_{\text{SR}}}{2\tau_R}\right)} C_{\text{Tank}}}. \tag{61}$$

Although more accurate than (58) and (60), this is still an approximation that loses accuracy when $T_{\text{SR}}/\tau_R < 2$ (or $T_{\text{SF}}/\tau_F < 2$).

Figure 6 and Figure 7 compare the derived analytical models for the stepwise driver ($E_{\text{Load-Driver}}$) from Section 3 of this work, Equation (58) based on [6], Equation (60) based on [7], and Equation (61) from combining [6] and [7] against the results of a circuit simulator. The figures show that the model presented in this work is the only analytical model that accurately takes into account finite $C_{Tank}$ and the step settling time $T_{SR}$ (or $T_{SF}$).

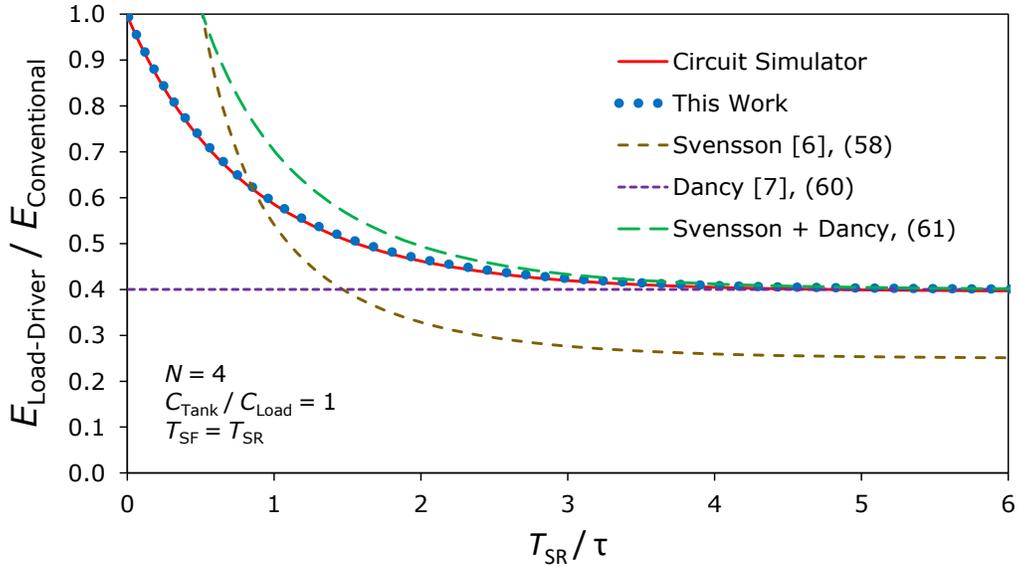

Figure 6. Comparing analytic model results of this work, Svensson [6], Dancy [7], and a combination of Svensson + Dancy against the results from a circuit simulator. Stepwise driver energy is normalized to that of a conventional driver ($N = 1$). Switch $S_R$ on-time $T_{SR}$ is normalized to the RC settling time constant, $\tau = \tau_R = \tau_F$. $T_{SF} = T_{SR}$. $N = 4$. $C_{Tank} / C_{Load} = 1$.

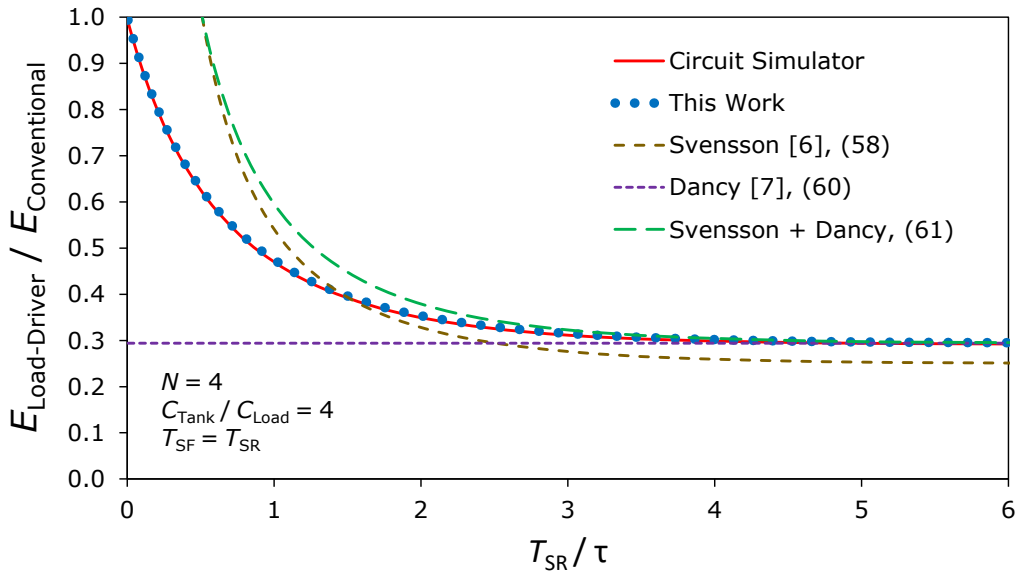

Figure 7. Comparing analytic model results of this work, Svensson [6], Dancy [7], and a combination of Svensson + Dancy against the results from a circuit simulator. Stepwise driver energy is normalized to that of a conventional driver ($N = 1$). Switch $S_R$ on-time $T_{SR}$ is normalized to the RC settling time constant, $\tau = \tau_R = \tau_F$. $T_{SF} = T_{SR}$. $N = 4$. $C_{Tank} / C_{Load} = 4$.

# 5 Conclusion

An accurate analytical model of a stepwise adiabatic driver has been presented. The model solves for all intermediate voltage levels (steps) for both rising and falling edges and calculates the energy consumed by the stepwise driver. The solution can be obtained by solving a matrix equation that can be built using the provided pseudocode. The accuracy of the model was compared with results from a circuit simulator as well as results from models from prior works. The model presented here provides the only closed-form solution that accurately takes into account the number of steps, step settling time, and the tank capacitance of the stepwise driver.